\pdfoutput=1
\documentclass[journal]{IEEEtran}
\usepackage{amsmath,amsfonts,graphicx,tabularx,color}
\usepackage{amssymb,textcomp,mathtools}
\usepackage{bm,upgreek,algorithm,hyperref}
\usepackage{multirow}
\usepackage{enumerate}
\usepackage{makecell,textcomp}
\usepackage{setspace, cite}
\RequirePackage{doi}

\def\thline{\noalign{\hrule height 1.0pt}}

\renewcommand{\vec}[1]{\bm{\mathrm{#1}}}

\title{Conv-TasNet: Surpassing Ideal Time-Frequency Magnitude Masking for Speech Separation}

\iftrue
\author{Yi~Luo, Nima~Mesgarani}
\fi
\begin{document}
\maketitle
\begin{abstract}
Single-channel, speaker-independent speech separation methods have recently seen great progress. However, the accuracy, latency, and computational cost of such methods remain insufficient. The majority of the previous methods have formulated the separation problem through the time-frequency representation of the mixed signal, which has several drawbacks, including the decoupling of the phase and magnitude of the signal, the suboptimality of time-frequency representation for speech separation, and the long latency in calculating the spectrograms. To address these shortcomings, we propose a fully-convolutional time-domain audio separation network (Conv-TasNet), a deep learning framework for end-to-end time-domain speech separation. Conv-TasNet uses a linear encoder to generate a representation of the speech waveform optimized for separating individual speakers. Speaker separation is achieved by applying a set of weighting functions (masks) to the encoder output. The modified encoder representations are then inverted back to the waveforms using a linear decoder. The masks are found using a temporal convolutional network (TCN) consisting of stacked 1-D dilated convolutional blocks, which allows the network to model the long-term dependencies of the speech signal while maintaining a small model size. The proposed Conv-TasNet system significantly outperforms previous time-frequency masking methods in separating two- and three-speaker mixtures. Additionally, Conv-TasNet surpasses several ideal time-frequency magnitude masks in two-speaker speech separation as evaluated by both objective distortion measures and subjective quality assessment by human listeners. Finally, Conv-TasNet has a significantly smaller model size and a shorter minimum latency, making it a suitable solution for both offline and real-time speech separation applications. This study therefore represents a major step toward the realization of speech separation systems for real-world speech processing technologies.
\end{abstract}
\begin{IEEEkeywords}
Source separation, single-channel, time-domain, deep learning, real-time
\end{IEEEkeywords}
\section{Introduction}
\label{sec:intro}
Robust speech processing in real-world acoustic environments often requires automatic speech separation. Because of the importance of this research topic for speech processing technologies, numerous methods have been proposed for solving this problem. However, the accuracy of speech separation, particularly for new speakers, remains inadequate.

Most previous speech separation approaches have been formulated in the time-frequency (T-F, or spectrogram) representation of the mixture signal, which is estimated from the waveform using the short-time Fourier transform (STFT) \cite{wang2018supervised}. Speech separation methods in the T-F domain aim to approximate the clean spectrogram of the individual sources from the mixture spectrogram. This process can be performed by directly approximating the spectrogram representation of each source from the mixture using nonlinear regression techniques, where the clean source spectrograms are used as the training target \cite{lu2013speech, xu2014experimental, xu2015regression}. Alternatively, a weighting function (mask) can be estimated for each source to multiply each T-F bin in the mixture spectrogram to recover the individual sources. In recent years, deep learning has greatly advanced the performance of time-frequency masking methods by increasing the accuracy of the mask estimation \cite{isik2016single, yu2017permutation, kolbaek2017multitalker, chen2017deep, luo2017speaker, wang2018alternative, wang2018end, li2018cbldnn}. In both the direct method and the mask estimation method, the waveform of each source is calculated using the inverse short-time Fourier transform (iSTFT) of the estimated magnitude spectrogram of each source together with either the original or the modified phase of the mixture sound.

While time-frequency masking remains the most commonly used method for speech separation, this method has several shortcomings. First, STFT is a generic signal transformation that is not necessarily optimal for speech separation. Second, accurate reconstruction of the phase of the clean sources is a nontrivial problem, and the erroneous estimation of the phase introduces an upper bound on the accuracy of the reconstructed audio. This issue is evident by the imperfect reconstruction accuracy of the sources even when the ideal clean magnitude spectrograms are applied to the mixture. Although methods for phase reconstruction can be applied to alleviate this issue \cite{griffin1984signal, le2008explicit, wang2018end}, the performance of the method remains suboptimal. Third, successful separation from the time-frequency representation requires a high-resolution frequency decomposition of the mixture signal, which requires a long temporal window for the calculation of STFT. This requirement increases the minimum latency of the system, which limits its applicability in real-time, low-latency applications such as in telecommunication and hearable devices. For example, the window length of STFT in most speech separation systems is at least 32 ms \cite{isik2016single, kolbaek2017multitalker, chen2017deep} and is even greater in music separation applications, which require an even higher resolution spectrogram (higher than 90 ms) \cite{luo2017deep, jansson2017singing}.

Because these issues arise from formulating the separation problem in the time-frequency domain, a logical approach is to avoid decoupling the magnitude and the phase of the sound by directly formulating the separation in the time domain. Previous studies have explored the feasibility of time-domain speech separation through methods such as independent component analysis (ICA) \cite{choi2005blind} and time-domain non-negative matrix factorization (NMF) \cite{yoshii2013beyond}. However, the performance of these systems has not been comparable with the performance of time-frequency approaches, particularly in terms of their ability to scale and generalize to large data. On the other hand, a few recent studies have explored deep learning for time-domain audio separation \cite{venkataramani2017end, stoller2018wave, luo2018tasnet}. The shared idea in all these systems is to replace the STFT step for feature extraction with a data-driven representation that is jointly optimized with an end-to-end training paradigm. These representations and their inverse transforms can be explicitly designed to replace STFT and iSTFT. Alternatively, feature extraction together with separation can be implicitly incorporated into the network architecture, for example by using an end-to-end convolutional neural network (CNN) \cite{fu2018end, pascual2017segan}. These methods are different in how they extract features from the waveform and in terms of the design of the separation module. In \cite{venkataramani2017end}, a convolutional encoder motivated by discrete cosine transform (DCT) is used as the front-end. The separation is then performed by passing the encoder features to a multilayer perceptron (MLP). The reconstruction of the waveforms is achieved by inverting the encoder operation. In \cite{stoller2018wave}, the separation is incorporated into a U-Net 1-D CNN architecture \cite{ronneberger2015u} without explicitly transforming the input into a spectrogram-like representation. However, the performance of these methods on a large speech corpus such as the benchmark introduced in \cite{hershey2016deep} has not been tested. Another such method is the time-domain audio separation network (TasNet) \cite{luo2018tasnet, luo2018real}. In TasNet, the mixture waveform is modeled with a convolutional encoder-decoder architecture, which consists of an encoder with a non-negativity constraint on its output and a linear decoder for inverting the encoder output back to the sound waveform. This framework is similar to the ICA method when a non-negative mixing matrix is used \cite{wang2010nonnegative} and to the semi-nonnegative matrix factorization method (semi-NMF) \cite{ding2010convex}, where the basis signals are the parameters of the decoder. The separation step in TasNet is done by finding a weighting function for each source (similar to time-frequency masking) for the encoder output at each time step. It has been shown that TasNet has achieved better or comparable performance with various previous T-F domain systems, showing its effectiveness and potential. 

While TasNet outperformed previous time-frequency speech separation methods in both causal and non-causal implementations, the use of a deep long short-term memory (LSTM) network as the separation module in the original TasNet significantly limited its applicability. First, choosing smaller kernel size (i.e. length of the waveform segments) in the encoder increases the length of the encoder output, which makes the training of the LSTMs unmanageable. Second, the large number of parameters in deep LSTM network significantly increases its computational cost and limits its applicability to low-resource, low-power platforms such as wearable hearing devices. The third problem which we will illustrate in this paper is caused by the long temporal dependencies of LSTM networks which often results in inconsistent separation accuracy, for example, when changing the starting point of the mixture. To alleviate the limitations of the previous TasNet, we propose the fully-convolutional TasNet (Conv-TasNet) that uses only convolutional layers in all stages of processing. Motivated by the success of temporal convolutional network (TCN) models \cite{lea2016temporal, lea2017temporal, bai2018empirical}, Conv-TasNet uses stacked dilated 1-D convolutional blocks to replace the deep LSTM networks for the separation step. The use of convolution allows parallel processing on consecutive frames or segments to greatly speed up the separation process and also significantly reduces the model size. To further decrease the number of parameters and the computational cost, we substitute the original convolution operation with depthwise separable convolution \cite{chollet2016xception, howard2017mobilenets}. We show that with these modifications, Conv-TasNet significantly increases the separation accuracy over the previous LSTM-TasNet in both causal and non-causal implementations. Moreover, the separation accuracy of Conv-TasNet surpasses the performance of ideal time-frequency magnitude masks, including the ideal binary mask (IBM \cite{wang2005ideal}), ideal ratio mask (IRM \cite{li2009optimality, wang2014training}), and Winener filter-like mask (WFM \cite{erdogan2015phase}) in both signal-to-distortion ratio (SDR) and subjective (mean opinion score, MOS) measures.

The rest of the paper is organized as follows. We introduce the proposed Conv-TasNet in section~\ref{sec:model}, describe the experimental procedures in section~\ref{sec:procedures}, and show the experimental results and analysis in section~\ref{sec:exp}. 

\section{Convolutional Time-domain Audio Separation Network}
\label{sec:model}
\begin{figure*}[!htp]
	\small
	\centering
	\includegraphics[width=2\columnwidth]{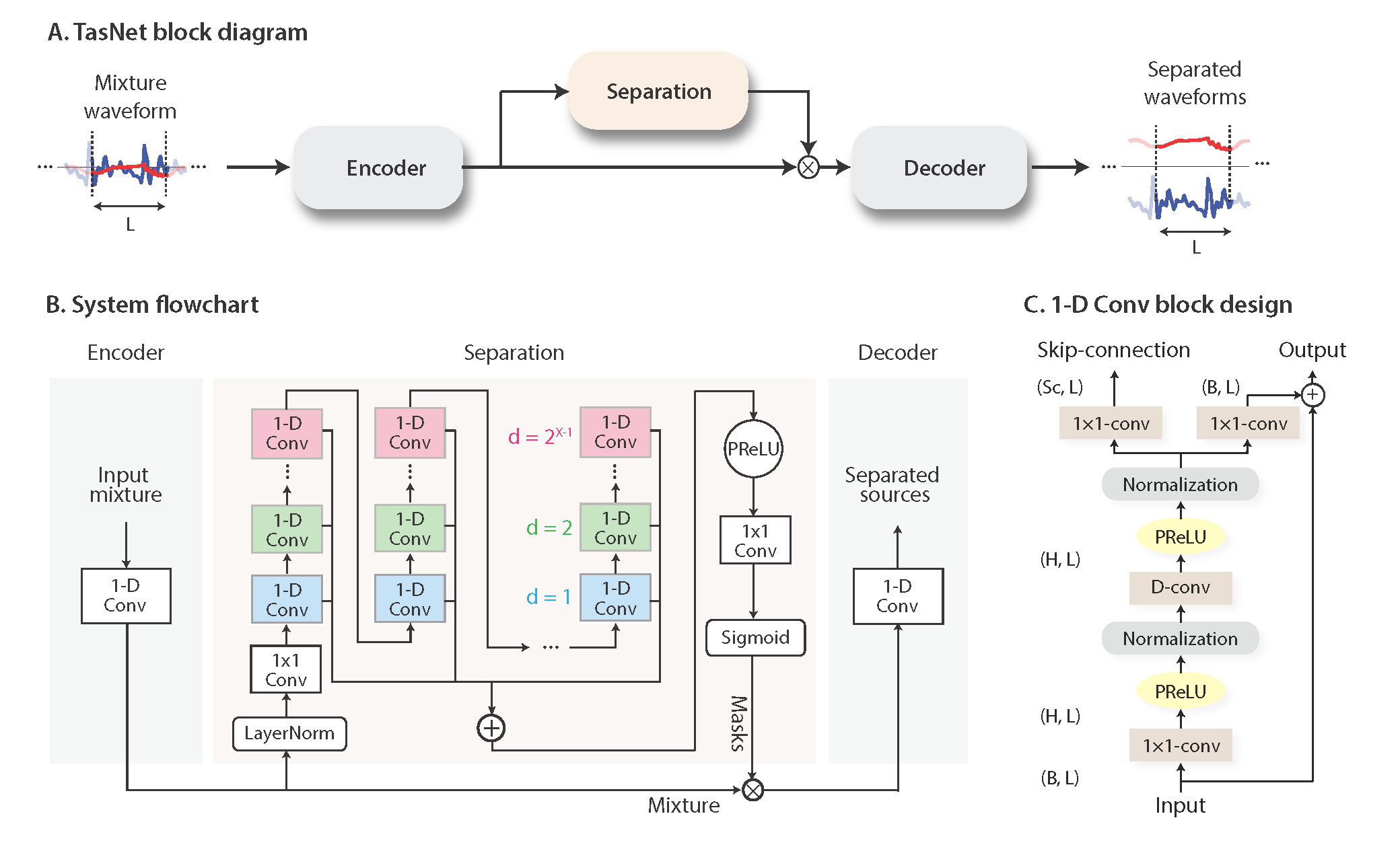}
	\caption{(A): the block diagram of the TasNet system. An encoder maps a segment of the mixture waveform to a high-dimensional representation and a separation module calculates a multiplicative function (i.e., a mask) for each of the target sources. A decoder reconstructs the source waveforms from the masked features. (B): A flowchart of the proposed system. A 1-D convolutional autoencoder models the waveforms and a temporal convolutional network (TCN) separation module estimates the masks based on the encoder output. Different colors in the 1-D convolutional blocks in TCN denote different dilation factors. (C): The design of 1-D convolutional block. Each block consists of a $1\times 1\text{-}conv$ operation followed by a depthwise convolution ($D-conv$) operation, with nonlinear activation function and normalization added between each two convolution operations. Two linear $1\times 1{-}conv$ blocks serve as the residual path and the skip-connection path respectively.}
	\label{fig:flowchart}
\end{figure*}

The fully-convolutional time-domain audio separation network (Conv-TasNet) consists of three processing stages, as shown in figure~\ref{fig:flowchart} (A): encoder, separation, and decoder. First, an encoder module is used to transform short segments of the mixture waveform into their corresponding representations in an intermediate feature space.
This representation is then used to estimate a multiplicative function (mask) for each source at each time step. The source waveforms are then reconstructed by transforming the masked encoder features using a decoder module. We describe the details of each stage in this section.

\subsection{Time-domain speech separation}

The problem of single-channel speech separation can be formulated in terms of estimating $C$ sources $s_1(t), \ldots, s_c(t) \in \mathbb{R}^{1\times T}$, given the discrete waveform of the mixture $x(t) \in \mathbb{R}^{1\times T}$, where
\begin{align}
x(t) = \sum_{i=1}^C s_i(t)
\label{eqn:prob}
\end{align}

In time-domain audio separation, we aim to directly estimate $s_i(t), i = 1, \ldots, C$, from $x(t)$. 

\subsection{Convolutional encoder-decoder}

The input mixture sound can be divided into overlapping segments of length $L$, represented by $\vec{x}_k \in \mathbb{R}^{1\times L}$, where $k = 1,\dots,\hat{T}$ denotes the segment index and $\hat{T}$ denotes the total number of segments in the input. $\vec{x}_k$ is transformed into a $N$-dimensional representation, $\vec{w} \in \mathbb{R}^{1\times N}$ by a 1-D convolution operation, which is reformulated as a matrix multiplication (the index $k$ is dropped from now on):
\begin{align}
\vec{w} = \mathcal{H}(\vec{x}\vec{U})
\label{eqn:enc}
\end{align}
where $\vec{U}\in \mathbb{R}^{N\times L}$ contains $N$ vectors (encoder basis functions) with length $L$ each, and $\mathcal{H}(\cdot)$ is an optional nonlinear function. In \cite{luo2018tasnet, luo2018real}, $\mathcal{H}(\cdot)$ was the rectified linear unit (ReLU) to ensure that the representation is non-negative. The decoder reconstructs the waveform from this representation using a 1-D transposed convolution operation, which can be reformulated as another matrix multiplication:
\begin{align}
\hat{\vec{x}} = \vec{w}\vec{V}
\label{eqn:dec}
\end{align}
where $\hat{\vec{x}} \in \mathbb{R}^{1\times L}$ is the reconstruction of $\vec{x}$, and the rows in $\vec{V} \in \mathbb{R}^{N\times L}$ are the decoder basis functions, each with length $L$. The overlapping reconstructed segments are summed together to generate the final waveforms.

Although we reformulate the encoder/decoder operations as matrix multiplication, the term "convolutional autoencoder" is used because in actual model implementation, convolutional and transposed convolutional layers can more easily handle the overlap between segments and thus enable faster training and better convergence. \footnote{With our Pytorch implementation, this is possibly due to the different \textit{autograd} mechanisms in fully-connected layer and 1-D (transposed) convolutional layers.}

\subsection{Estimating the separation masks}
\label{sec:mask}

The separation for each frame is performed by estimating $C$ vectors (masks) $\vec{m}_{i} \in \mathbb{R}^{1\times N}, i=1, \ldots, C$ where $C$ is the number of speakers in the mixture that is multiplied by the encoder output $\vec{w}$. The mask vectors $\vec{m}_{i}$ have the constraint that $\vec{m}_{i} \in [0, 1]$. The representation of each source, $\vec{d}_i \in \mathbb{R}^{1\times N}$, is then calculated by applying the corresponding mask, $\vec{m}_{i}$, to the mixture representation $\vec{w}$:
\begin{gather}
\vec{d}_i = \vec{w} \odot \vec{m}_i
\label{eqn:mask}
\end{gather}
where $\odot$ denotes element-wise multiplication. The waveform of each source $\hat{\vec{s}}_i, i=1,\ldots,C$ is then reconstructed by the decoder:
\begin{align}
\hat{\vec{s}}_i =\vec{d}_i\vec{V}
\label{eqn:dec_source}
\end{align}
The unit summation constraint in \cite{luo2018tasnet, luo2018real}, $\sum_{i=1}^C \vec{m}_{i} = \mathbb{\vec{1}}$, was applied based on the assumption that the encoder-encoder architecture can perfectly reconstruct the input mixture. In section~\ref{sec:exp-design}, we will examine the consequence of relaxing this unity summation constraint on separation accuracy.

\subsection{Convolutional separation module}

Motivated by the temporal convolutional network (TCN) \cite{lea2016temporal, lea2017temporal, bai2018empirical}, we propose a fully-convolutional separation module that consists of stacked 1-D dilated convolutional blocks, as shown in figure~\ref{fig:flowchart} (B). TCN was proposed as a replacement for RNNs in various sequence modeling tasks. Each layer in a TCN consists of 1-D convolutional blocks with increasing dilation factors. The dilation factors increase exponentially to ensure a sufficiently large temporal context window to take advantage of the long-range dependencies of the speech signal, as denoted with different colors in figure~\ref{fig:flowchart} (B). In Conv-TasNet, $M$ convolutional blocks with dilation factors $1, 2, 4, \ldots, 2^{M-1}$ are repeated $R$ times. The input to each block is zero padded accordingly to ensure the output length is the same as the input. The output of the TCN is passed to a convolutional block with kernel size $1$ ($1\times 1{-}conv$ block, also known as \textit{pointwise} convolution) for mask estimation. The $1\times 1{-}conv$ block together with a nonlinear activation function estimates $C$ mask vectors for the $C$ target sources.

Figure~\ref{fig:flowchart} (C) shows the design of each 1-D convolutional block. The design of the 1-D convolutional blocks follows \cite{van2016wavenet}, where a residual path and a skip-connection path are applied: the residual path of a block serves as the input to the next block, and the skip-connection paths for all blocks are summed up and used as the output of the TCN. To further decrease the number of parameters, depthwise separable convolution ($S\text{-}conv(\cdot)$) is used to replace standard convolution in each convolutional block. Depthwise separable convolution (also referred to as separable convolution) has proven effective in image processing tasks \cite{chollet2016xception, howard2017mobilenets} and neural machine translation tasks \cite{kaiser2017depthwise}. The depthwise separable convolution operator decouples the standard convolution operation into two consecutive operations, a depthwise convolution ($D\text{-}conv(\cdot)$) followed by pointwise convolution ($1\times 1{-}conv(\cdot)$):
\begin{align}
D\text{-}conv(\vec{Y}, \vec{K}) &= concat(\vec{y}_j \circledast \vec{k}_j), j=1, \ldots, N \\
S\text{-}conv(\vec{Y},\vec{K},\vec{L}) &= D\text{-}conv(\vec{Y}, \vec{K}) \circledast \vec{L}
\label{eqn:depthwise}
\end{align}
where $\vec{Y} \in \mathbb{R}^{G\times M}$ is the input to $S\text{-}conv(\cdot)$, $\vec{K} \in \mathbb{R}^{G\times P}$ is the convolution kernel with size $P$, $\vec{y}_j \in \mathbb{R}^{1\times M}$ and $\vec{k}_j \in \mathbb{R}^{1\times P}$ are the rows of matrices $\vec{Y}$ and $\vec{K}$, respectively, $\vec{L} \in \mathbb{R}^{G\times H\times 1}$ is the convolution kernel with size 1, and $\circledast$ denotes the convolution operation. In other words, the $D\text{-}conv(\cdot)$ operation convolves each row of the input $Y$ with the corresponding row of matrix $K$, and the $1\times 1{-}conv$ block linearly transforms the feature space. In comparison with the standard convolution with kernel size $\hat{\vec{K}} \in \mathbb{R}^{G\times H\times P}$, depthwise separable convolution only contains $G\times P + G\times H$ parameters, which decreases the model size by a factor of $\frac{H\times P}{H+P}\approx P$ when $H \gg P$. 

A nonlinear activation function and a normalization operation are added after both the first $1\times 1\text{-}conv$ and $D\text{-}conv$ blocks respectively. The nonlinear activation function is the parametric rectified linear unit (PReLU) \cite{he2015delving}:
\begin{align}
PReLU(x) = 
\begin{cases}x, \quad \text{if} \, x \geq 0\\
\alpha x, \quad \text{otherwise}
\end{cases}
\label{eqn:prelu}
\end{align}
where $\alpha \in \mathbb{R}$ is a trainable scalar controlling the negative slope of the rectifier. The choice of the normalization method in the network depends on the causality requirement. For noncausal configuration, we found empirically that global layer normalization (gLN) outperforms all other normalization methods. In gLN, the feature is normalized over both the channel and the time dimensions:
\begin{align}
gLN(\vec{F}) &= \frac{\vec{F} - E[\vec{F}]}{\sqrt{Var[\vec{F}] + \epsilon}} \odot \gamma + \beta \\
E[\vec{F}] &= \frac{1}{NT} \sum_{NT}{\vec{F}} \\
Var[\vec{F}] &= \frac{1}{NT} \sum_{NT}{(\vec{F} - E[\vec{F}])^2}
\label{eqn:gLN}
\end{align}
where $\vec{F} \in \mathbb{R}^{N\times T}$ is the feature, $\gamma, \beta \in \mathbb{R}^{N\times 1}$ are trainable parameters, and $\epsilon$ is a small constant for numerical stability. This is identical to the standard layer normalization applied in computer vision models where the channel and time dimension correspond to the width and height dimension in an image \cite{ba2016layer}. In causal configuration, gLN cannot be applied since it relies on the future values of the signal at any time step. Instead, we designed a cumulative layer normalization (cLN) operation to perform step-wise normalization in the causal system:
\begin{align}
cLN(\vec{f}_k) &= \frac{\vec{f}_k - E[\vec{f}_{t\leq k}]}{\sqrt{Var[\vec{f}_{t\leq k}] + \epsilon}} \odot \gamma + \beta \\
E[\vec{f}_{t\leq k}] &= \frac{1}{Nk} \sum_{Nk}{\vec{f}_{t\leq k}} \\
Var[\vec{f}_{t\leq k}] &= \frac{1}{Nk} \sum_{Nk}{(\vec{f}_{t\leq k} - E[\vec{f}_{t\leq k}])^2}
\label{eqn:cLN}
\end{align}
where $\vec{f}_k \in \mathbb{R}^{N\times 1}$ is the $k$-th frame of the entire feature $\vec{F}$, $\vec{f}_{t\leq k} \in \mathbb{R}^{N\times k}$ corresponds to the feature of $k$ frames $[\vec{f}_1, \ldots, \vec{f}_k]$, and $\gamma, \beta \in \mathbb{R}^{N\times 1}$ are trainable parameters applied to all frames. To ensure that the separation module is invariant to the scaling of the input, the selected normalization method is applied to the encoder output $\vec{w}$ before it is passed to the separation module.

At the beginning of the separation module, a linear $1\times 1\text{-}conv$ block is added as a bottleneck layer. This block determines the number of channels in the input and residual path of the subsequent convolutional blocks. For instance, if the linear bottleneck layer has $B$ channels, then for a 1-D convolutional block with $H$ channels and kernel size $P$, the size of the kernel in the first $1\times 1\text{-}conv$ block and the first $D\text{-}conv$ block should be $\vec{O} \in \mathbb{R}^{B\times H\times 1}$ and $\vec{K} \in \mathbb{R}^{H\times P}$ respectively, and the size of the kernel in the residual paths should be $\vec{L}_{Rs} \in \mathbb{R}^{H\times B\times 1}$. The number of output channels in the skip-connection path can be different than $B$, and we denote the size of kernels in that path as $\vec{L}_{Sc} \in \mathbb{R}^{H\times Sc\times 1}$.

\section{Experimental procedures}
\label{sec:procedures}
\begin{figure*}[!htp]
	\small
	\centering
	\includegraphics[width=1.8\columnwidth]{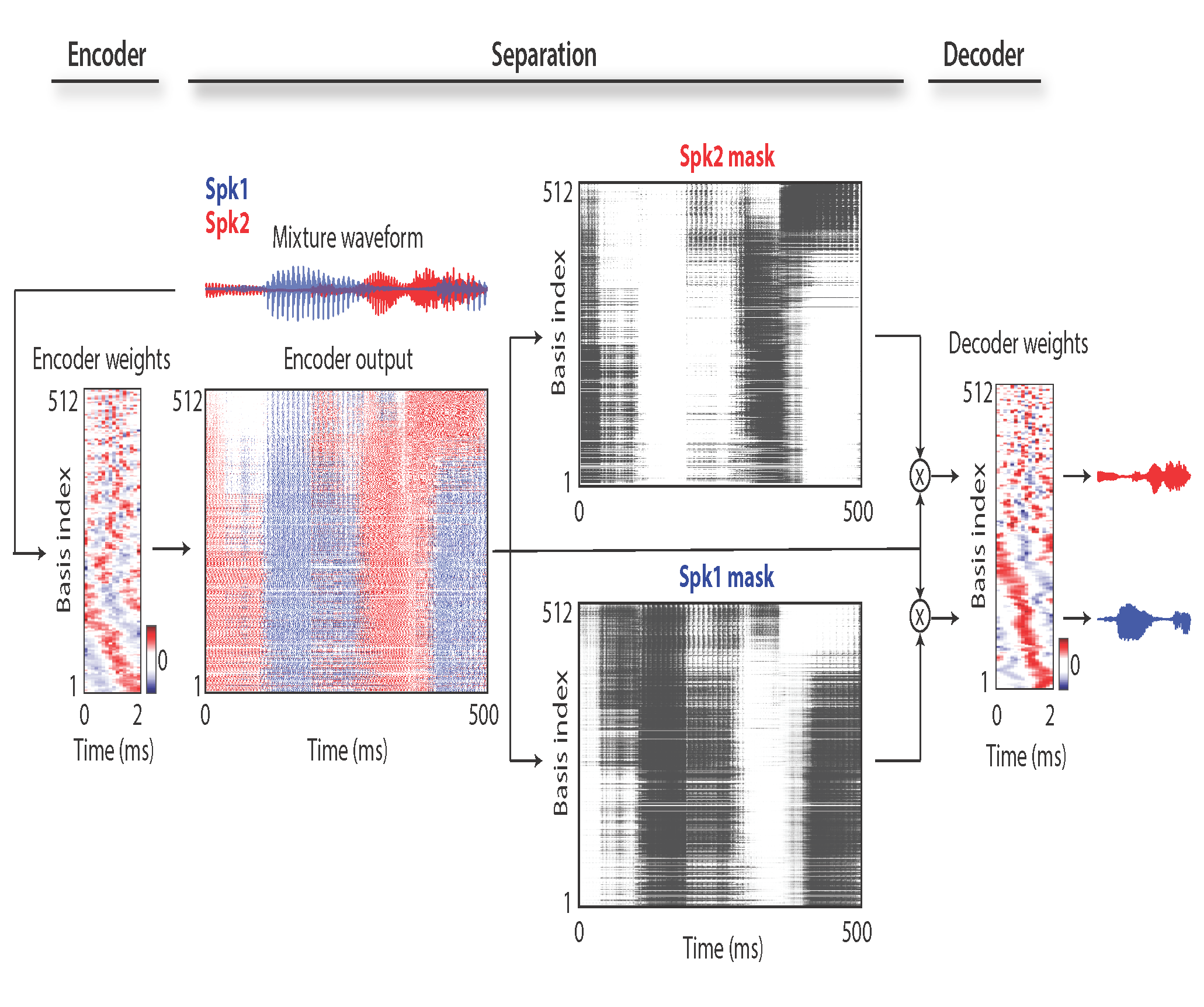}
	\caption{Visualization of the encoder and decoder basis functions, encoder representation, and source masks for a sample 2-speaker mixture. The speakers are shown in red and blue. The encoder representation is colored according to the power of each speaker at each basis function and point in time. The basis functions are sorted according to their Euclidean similarity and show diversity in frequency and phase tuning.}
	\label{fig:sample}
\end{figure*}

\subsection{Dataset}

We evaluated our system on two-speaker and three-speaker speech separation problems using the WSJ0-2mix and WSJ0-3mix datasets \cite{hershey2016deep}. 30 hours of training and 10 hours of validation data are generated from speakers in si\_tr\_s from the datasets. The speech mixtures are generated by randomly selecting utterances from different speakers in the Wall Street Journal dataset (WSJ0) and mixing them at random signal-to-noise ratios (SNR) between -5 dB and 5 dB. 5 hours of evaluation set is generated in the same way using utterances from 16 unseen speakers in si\_dt\_05 and si\_et\_05. The scripts for creating the dataset can be found at \cite{web2017deep}. All the waveforms are resampled at 8 kHz.

\subsection{Experiment configurations}
\label{sec:config}

The networks are trained for 100 epochs on 4-second long segments. The initial learning rate is set to $1e^{-3}$. The learning rate is halved if the accuracy of validation set is not improved in 3 consecutive epochs. Adam \cite{kingma2014adam} is used as the optimizer. A 50\% stride size is used in the convolutional autoencoder (i.e. 50\% overlap between consecutive frames). Gradient clipping with maximum $L_2$-norm of 5 is applied during training. The hyperparameters of the network are shown in table~\ref{tab:parameter}. A Pytorch implementation of the Conv-TasNet model can be found at \footnote{https://github.com/naplab/Conv-TasNet}.

\begin{table}[!htbp]
	\small
	\centering
	\caption{Hyperparameters of the network.}
	\vspace{0.2cm}
	\label{tab:parameter}
	\begin{tabular}{c|c}
		\thline
		Symbol & Description\\
		\thline
		$N$ & \thead{Number of filters in autoencoder}\\
		\hline
		$L$ & \thead{Length of the filters (in samples)}\\
		\hline
		$B$ & \thead{Number of channels in bottleneck\\ and the residual paths' $1\times 1\text{-}conv$ blocks} \\
		\hline
		$Sc$ & \thead{Number of channels in skip-connection\\ paths' $1\times 1\text{-}conv$ blocks} \\
		\hline
		$H$ & \thead{Number of channels in convolutional blocks}\\
		\hline
		$P$ & \thead{Kernel size in convolutional blocks}\\
		\hline
		$X$ & \thead{Number of convolutional blocks in each repeat}\\
		\hline
		$R$ & \thead{Number of repeats}\\
		\thline
	\end{tabular}
\end{table}

\subsection{Training objective}

The objective of training the end-to-end system is maximizing the scale-invariant source-to-noise ratio (SI-SNR), which has commonly been used as the evaluation metric for source separation replacing the standard source-to-distortion ratio (SDR) \cite{vincent2006performance, isik2016single, luo2017speaker}. SI-SNR is defined as:
\begin{align}
\begin{cases}
\vec{s}_{target} := \frac{\langle \hat{\vec{s}}, \vec{s} \rangle \vec{s}}{\left \| \vec{s} \right \|^2}\\
\vec{e}_{noise} := \hat{\vec{s}} - \vec{s}_{target}\\
\text{SI-SNR} := 10\,log_{10}\frac{\left \| \vec{s}_{target} \right \|^2}{\left \| \vec{e}_{noise} \right \|^2}
\end{cases}
\label{eqn:si-snr}
\end{align}
where $\hat{\vec{s}} \in \mathbb{R}^{1\times T}$ and $\vec{s} \in \mathbb{R}^{1\times T}$ are the estimated and original clean sources, respectively, and $\left \| \vec{s} \right \|^2 = \langle \vec{s}, \vec{s} \rangle$ denotes the signal power. Scale invariance is ensured by normalizing $\hat{\vec{s}}$ and $\vec{s}$ to zero-mean prior to the calculation. Utterance-level permutation invariant training (uPIT) is applied during training to address the source permutation problem \cite{kolbaek2017multitalker}.

\subsection{Evaluation metrics}

We report the scale-invariant signal-to-noise ratio improvement (SI-SNRi) and signal-to-distortion ratio improvement (SDRi) \cite{vincent2006performance} as objective measures of separation accuracy. SI-SNR is defined in equation~\ref{eqn:si-snr}. The reported improvements in tables~\ref{tab:result-system} to~\ref{tab:result-comp3} indicate the additive values over the original mixture. In addition to the distortion metrics, we also evaluated the quality of the separated mixtures using both the perceptual evaluation of subjective quality (PESQ, \cite{rix2001perceptual}) and the mean opinion score (MOS) \cite{MOS} by asking 40 normal hearing subjects to rate the quality of the separated mixtures. All human testing procedures were approved by the local institutional review board (IRB) at Columbia University in the City of New York. 

\subsection {Comparison with ideal time-frequency masks}

Following the common configurations in \cite{isik2016single, luo2017speaker, kolbaek2017multitalker}, the ideal time-frequency masks were calculated using STFT with a 32 ms window size and 8 ms hop size with a Hanning window. The ideal masks include the ideal binary mask (IBM), ideal ratio mask (IRM), and Wiener filter-like mask (WFM), which are defined for source $i$ as:
\begin{align}
IBM_i(f,t) &= 
\begin{cases}
1, \, \left |\mathcal{S}_i(f,t)\right | > \left |\mathcal{S}_{j\neq i}(f,t)\right | \\
0, \, \text{otherwise}
\end{cases}\\
IRM_i(f,t) &= \frac{\left |\mathcal{S}_i(f,t)\right |}{\sum_{j=1}^C \left |\mathcal{S}_j(f,t)\right |} \\
WFM_i(f,t) &= \frac{\left |\mathcal{S}_i(f,t)\right |^2}{\sum_{j=1}^C \left |\mathcal{S}_j(f,t)\right |^2}
\label{eqn:real-masks}
\end{align}

where $\mathcal{S}_i(f,t) \in \mathbb{C}^{F\times T}$ are the complex-valued spectrograms of clean sources $i=1,\ldots, C$.

\section{Results}
\label{sec:exp}
\begin{table*}[!htbp]
	\small
	\centering
	\caption{The effect of different configurations in Conv-TasNet.}
	\vspace{0.2cm}
	\label{tab:result-sep2}
	\begin{tabular}{c|c|c|c|c|c|c|c|c|c|c|c|c|c}
		\thline
		\thead{$N$} & \thead{$L$} & \thead{$B$} & \thead{$H$} & \thead{$Sc$} & \thead{$P$} & \thead{$X$} & \thead{$R$} & \thead{Normali-\\zation} & \thead{Causal} & \thead{Receptive \\ field (s)} & \thead{Model \\ size} & \thead{SI-SNRi\\ (dB)} & \thead{SDRi\\ (dB)} \\
		\hline
		128 & 40 & 128 & 256 & 128 & 3 & 7 & 2 & gLN & \texttimes & 1.28 & 1.5M & 13.0 & 13.3 \\
		256 & 40 & 128 & 256 & 128 & 3 & 7 & 2 & gLN & \texttimes & 1.28 & 1.5M & 13.1 & 13.4 \\
		512 & 40 & 128 & 256 & 128 & 3 & 7 & 2 & gLN & \texttimes & 1.28 & 1.7M & 13.3 & 13.6\\
		\hline
		512 & 40 & 128 & 256 & 256 & 3 & 7 & 2 & gLN & \texttimes & 1.28 & 2.4M & 13.0 & 13.3 \\ 
		512 & 40 & 128 & 512 & 128 & 3 & 7 & 2 & gLN & \texttimes & 1.28 & 3.1M & 13.3 & 13.6 \\
		512 & 40 & 128 & 512 & 512 & 3 & 7 & 2 & gLN & \texttimes & 1.28 & 6.2M & 13.5 & 13.8 \\
		512 & 40 & 256 & 256 & 256 & 3 & 7 & 2 & gLN & \texttimes & 1.28 & 3.2M & 13.0 & 13.3 \\
		512 & 40 & 256 & 512 & 256 & 3 & 7 & 2 & gLN & \texttimes & 1.28 & 6.0M & 13.4 & 13.7 \\
		512 & 40 & 256 & 512 & 512 & 3 & 7 & 2 & gLN & \texttimes & 1.28 & 8.1M & 13.2 & 13.5 \\
		\hline
		512 & 40 & 128 & 512 & 128 & 3 & 6 & 4 & gLN & \texttimes & 1.27 & 5.1M & 14.1 & 14.4 \\
		512 & 40 & 128 & 512 & 128 & 3 & 4 & 6 & gLN & \texttimes & 0.46 & 5.1M & 13.9 & 14.2 \\
		512 & 40 & 128 & 512 & 128 & 3 & 8 & 3 & gLN & \texttimes & 3.83 & 5.1M & 14.5 & 14.8 \\
		\hline
		512 & 32 & 128 & 512 & 128 & 3 & 8 & 3 & gLN & \texttimes & 3.06 & 5.1M & 14.7 & 15.0 \\
		512 & 16 & 128 & 512 & 128 & 3 & 8 & 3 & gLN & \texttimes & 1.53 & 5.1M & \bf{15.3} & \bf{15.6} \\
		512 & 16 & 128 & 512 & 128 & 3 & 8 & 3 & cLN & \checkmark & 1.53 & 5.1M & 10.6 & 11.0 \\
		\thline
	\end{tabular}
\end{table*}

Figure~\ref{fig:sample} visualizes all the internal variables of Conv-TasNet for one example mixture sound with two overlapping speakers (denoted by red and blue). The encoder and decoder basis functions are sorted by the similarity of the Euclidean distance of the basis functions found using the unweighted pair group method with arithmetic mean (UPGMA) method \cite{sokal1958statistical}. The basis functions show a diversity of frequency and phase tuning. The representation of the encoder is colored according to the power of each speaker at the corresponding basis output at each time point, demonstrating the sparsity of the encoder representation. As can be seen in figure~\ref{fig:sample}, the estimated masks for the two speakers highly resemble their encoder representations, which allows for the suppression of the encoder outputs that correspond to the interfering speaker and the extraction of the target speaker in each mask. The separated waveforms for the two speakers are estimated by the linear decoder, whose basis functions are shown in figure~\ref{fig:sample}. The separated waveforms are shown on the right.  

\subsection{Non-negativity of the encoder output}
\label{sec:exp-design}
The non-negativity of the encoder output was enforced in \cite{luo2018tasnet, luo2018real} using a rectified-linear nonlinearity (ReLU) function. This constraint was based on the assumption that the masking operation on the encoder output is only meaningful when the mixture and speaker waveforms can be represented with a non-negative combination of the basis functions, since an unbounded encoder representation may result in unbounded masks. However, by removing the nonlinear function $\mathcal{H}$, another assumption can be made: with an unbounded but highly overcomplete representation of the mixture, a set of non-negative masks can still be found to reconstruct the clean sources. In this case, the overcompleteness of the representation is crucial. If there exist only a unique weight feature for the mixture as well as for the sources, the non-negativity of the mask cannot be guaranteed. Also note that in both assumptions, we put no constraint on the relationship between the encoder and decoder basis functions $\vec{U}$ and $\vec{V}$, meaning that they are not forced to reconstruct the mixture signal perfectly. One way to explicitly ensure the autoencoder property is by choosing $\vec{V}$ to be the pseudo-inverse of $\vec{U}$ (i.e. least square reconstruction). The choice of encoder/decoder design affects the mask estimation: in the case of an autoencoder, the unit summation constraint must be satisfied; otherwise, the unit summation constraint is not strictly required. To illustrate this point, we compared five different encoder-decoder configurations: 
\begin{enumerate}
    \item Linear encoder with its pseudo-inverse (Pinv) as decoder, i.e. $\vec{w}=\vec{x}(\vec{V}^T\vec{V})^{-1}\vec{V}^T$ and $\hat{\vec{x}}=\vec{w}\vec{V}$, with Softmax function for mask estimation.
    \item Linear encoder and decoder where $\vec{w}=\vec{x}\vec{U}$ and $\hat{\vec{x}}=\vec{w}\vec{V}$, with Softmax or Sigmoid function for mask estimation.
    \item Encoder with ReLU activation and linear decoder where $\vec{w}=ReLU(\vec{x}\vec{U})$ and $\hat{\vec{x}}=\vec{w}\vec{V}$, with Softmax or Sigmoid function for mask estimation.
\end{enumerate}
Separation accuracy of different configurations in table~\ref{tab:result-system} shows that pseudo-inverse autoencoder leads to the worst performance, indicating that an explicit autoencoder configuration does not necessarily improve the separation score in this framework. The performance of all other configurations is comparable. Because linear encoder and decoder with Sigmoid function achieves a slightly better accuracy over other methods, we used this configuration in all the following experiments.

\begin{table}[!htbp]
	\small
	\centering
	\caption{Separation score for different system configurations.}
	\vspace{0.2cm}
	\label{tab:result-system}
	\begin{tabular}{c|c|c|c|c}
		\thline
		\thead{Encoder} & \thead{Mask} & \thead{Model \\ size} & \thead{SI-SNRi\\ (dB)} & \thead{SDRi\\ (dB)} \\
		\hline
		Pinv & Softmax & \multirow{5}{*}{1.5M} & 12.1 & 12.4 \\
		\multirow{2}{*}{Linear} & Softmax & & 12.9 & 13.2 \\
		 & Sigmoid & & \bf{13.1} & \bf{13.4} \\
		\multirow{2}{*}{ReLU} & Softmax & & 13.0 & 13.3 \\
		 & Sigmoid & & 12.9 & 13.2 \\
		\thline
	\end{tabular}
\end{table}

\subsection{Optimizing the network parameters}
\label{sec:param}

We evaluate the performance of Conv-TasNet on two speaker separation tasks as a function of different network parameters. Table~\ref{tab:result-sep2} shows the performance of the systems with different parameters, from which we can conclude the following statements:

\begin{enumerate}[(i)]
    \item Encoder/decoder: Increasing the number of basis signals in the encoder/decoder increases the overcompleteness of the basis signals and improves the performance.
    \item Hyperparameters in the 1-D convolutional blocks: A possible configuration consists of a small bottleneck size $B$ and a large number of channels in the convolutional blocks $H$. This matches the observation in \cite{sandler2018mobilenetv2}, where the ratio between the convolutional block and the bottleneck $H/B$ was found to be best around 5. Increasing the number of channels in the skip-connection block improves the performance while greatly increases the model size. Therefore, we selected a small skip-connection block as a trade-off between performance and model size.
    \item Number of 1-D convolutional blocks: When the receptive field is the same, deeper networks lead to better performance, possibly due to the increased model capacity.
    \item Size of receptive field: Increasing the size of receptive field leads to better performance, which shows the importance of modeling the temporal dependencies in the speech signal.
    \item Length of each segment: Shorter segment length consistently improves performance. Note that the best system uses a filter length of only 2 ms ($\frac{L}{fs}=\frac{16}{8000}=0.002s$), which makes it very difficult to train a deep LSTM network with the same $L$ due to the large number of time steps in the encoder output.
    \item Causality: Using a causal configuration leads to a significant drop in the performance. This drop could be due to the causal convolution and/or the layer normalization operations.
\end{enumerate}

\subsection{Comparison of Conv-TasNet with previous methods}
\label{sec:exp-compare}

We compared the separation accuracy of Conv-TasNet with previous methods using SDRi and SI-SNRi. Table~\ref{tab:result-comp2} compares the performance of Conv-TasNet with other state-of-the-art methods on the same WSJ0-2mix dataset. For all systems, we list the best results that have been reported in the literature. The numbers of parameters in different methods are based on our implementations, except for \cite{li2018cbldnn} which is provided by the authors. The missing values in the table are either because the numbers were not reported in the study or because the results were calculated with a different STFT configuration. The previous TasNet in \cite{luo2018real} is denoted by the (B)LSTM-TasNet. While the BLSTM-TasNet already outperformed IRM and IBM, the non-causal Conv-TasNet significantly surpasses the performance of all three ideal T-F masks in SI-SNRi and SDRi metrics with a significantly smaller model size comparing with all previous methods.

Table~\ref{tab:result-comp3} compares the performance of Conv-TasNet with those of other systems on a three-speaker speech separation task involving the WSJ0-3mix dataset. The non-causal Conv-TasNet system significantly outperforms all previous STFT-based systems in SDRi. While there is no prior result on a causal algorithm for three-speaker separation, the causal Conv-TasNet significantly outperforms even the other two non-causal STFT-based systems \cite{isik2016single, kolbaek2017multitalker}. Examples of separated audio for two and three speaker mixtures from both causal and non-causal implementations of Conv-TasNet are available online \cite{web2018tasnet}. 

\begin{table}[!htbp]
	\small
	\centering
	\caption{Comparison with other methods on WSJ0-2mix dataset.}
	\vspace{0.2cm}
	\label{tab:result-comp2}
	\begin{tabular}{c|c|c|c|c}
		\thline
		\thead{Method} & \thead{Model \\ size} & \thead{Causal} & \thead{SI-SNRi\\ (dB)} & \thead{SDRi\\ (dB)} \\
		\thline
		DPCL++ \cite{isik2016single} & 13.6M & \texttimes & 10.8 & -- \\
		uPIT-BLSTM-ST \cite{kolbaek2017multitalker} & 92.7M & \texttimes  & -- & 10.0 \\
		DANet \cite{chen2017deep} & 9.1M & \texttimes & 10.5 & --  \\
		ADANet \cite{luo2017speaker} & 9.1M & \texttimes & 10.4 & 10.8 \\
		cuPIT-Grid-RD \cite{xu2018single} & 47.2M & \texttimes & -- & 10.2 \\
		CBLDNN-GAT\cite{li2018cbldnn} & 39.5M & \texttimes & -- & 11.0 \\
		Chimera++ \cite{wang2018alternative} & 32.9M & \texttimes & 11.5 & 12.0 \\
		WA-MISI-5 \cite{wang2018end} & 32.9M & \texttimes & 12.6 & 13.1 \\
		BLSTM-TasNet \cite{luo2018real} & 23.6M & \texttimes & 13.2 & 13.6 \\
		\bf{Conv-TasNet-gLN} & \bf{5.1M} & \texttimes  & \bf{15.3} & \bf{15.6} \\
		\hline
		uPIT-LSTM \cite{kolbaek2017multitalker} & 46.3M & \checkmark  & -- & 7.0 \\
		LSTM-TasNet \cite{luo2018real} & 32.0M & \checkmark & \bf{10.8} & \bf{11.2} \\
		\bf{Conv-TasNet-cLN} & \bf{5.1M} & \checkmark & 10.6 & 11.0 \\
		\thline	
		IRM & -- &  --  & 12.2 & 12.6 \\
		IBM & -- &  --  & 13.0 & 13.5 \\
		WFM & -- &  --  & 13.4 & 13.8 \\
		\thline
	\end{tabular}
\end{table}

\begin{table}[!htbp]
	\small
	\centering
	\caption{Comparison with other systems on WSJ0-3mix dataset.}
	\vspace{0.2cm}
	\label{tab:result-comp3}
	\begin{tabular}{c|c|c|c|c}
		\thline
		\thead{Method} & \thead{Model \\ size} & \thead{Causal} & \thead{SI-SNRi\\ (dB)} &\thead{SDRi\\ (dB)} \\
		\thline
		
		DPCL++ \cite{isik2016single} & 13.6M & \texttimes & 7.1 & -- \\
		uPIT-BLSTM-ST \cite{kolbaek2017multitalker} & 92.7M & \texttimes & -- & 7.7 \\
		DANet \cite{chen2017deep} & 9.1M & \texttimes & 8.6 & 8.9 \\
		ADANet \cite{luo2017speaker} & 9.1M  & \texttimes & 9.1 & 9.4 \\
		\bf{Conv-TasNet-gLN} & \bf{5.1M} & \texttimes & \bf{12.7} & \bf{13.1} \\
		\hline
		\bf{Conv-TasNet-cLN} & \bf{5.1M} & \checkmark & \bf{7.8} & \bf{8.2}  \\
		\thline
		IRM & --  & -- & 12.5 & 13.0 \\
		IBM & --  & -- & 13.2 & 13.6 \\
		WFM & --  & -- & 13.6 & 14.0 \\
		\thline
	\end{tabular}
\end{table}

\subsection{Subjective and objective quality evaluation of Conv-TasNet}

In addition to SDRi and SI-SNRi, we evaluated the subjective and objective quality of the separated speech and compared with three ideal time-frequency magnitude masks. Table~\ref{tab:sub-masks} shows the PESQ score for Conv-TasNet and IRM, IBM, and WFM, where IRM has the highest score for both WSJ0-2mix and WSJ0-3mix dataset. However, since PESQ aims to predict the subjective quality of speech, human quality evaluation can be considered as the ground truth. Therefore, we conducted a psychophysics experiment in which we asked 40 normal hearing subjects to listen and rate the quality of the separated speech sounds. Because of the practical limitations of human psychophysics experiments, we restricted the subjective comparison of Conv-TasNet to the ideal ratio mask (IRM) which has the highest PESQ score among the three ideal masks (table~\ref{tab:sub-masks}). We randomly chose 25 two-speaker mixture sounds from the two-speaker test set (WSJ0-2mix). We avoided a possible selection bias by ensuring that the average PESQ scores for the IRM and Conv-TasNet separated sounds for the selected 25 samples were equal to the average PESQ scores over the entire test set (comparison of tables~\ref{tab:sub-masks} and~\ref{tab:sub-comp}). The length of each utterance was constrained to be within 0.5 standard deviation of the mean of the entire test set. The subjects were asked to rate the quality of the clean utterances, the IRM-separated utterances, and the Conv-TasNet separated utterances on the scale of 1 to 5 (1: bad, 2: poor, 3: fair, 4: good, 5: excellent). A clean utterance was first given as the reference for the highest possible score (i.e. 5). Then the clean, IRM, and Conv-TasNet samples were presented to the subjects in random order. The mean opinion score (MOS) of each of the 25 utterances was then averaged over the 40 subjects. 

\begin{figure*}[!htbp]
	\small
	\centering
	\includegraphics[width=1.7\columnwidth]{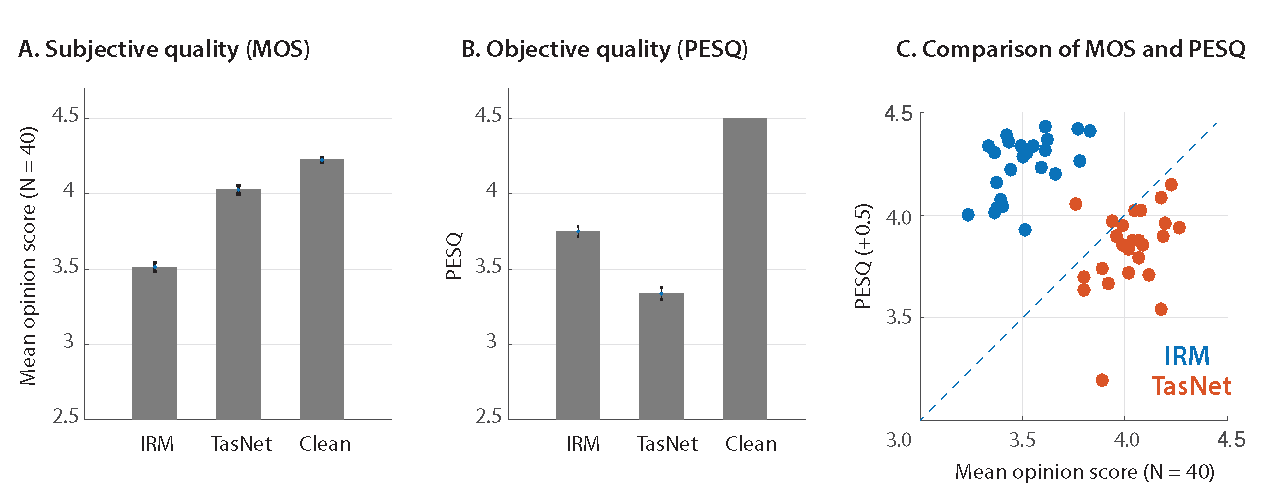}
	\caption{Subjective and objective quality evaluation of separated utterances in WSJ0-2mix. (A): The mean opinion scores (MOS, N = 40) for IRM, Conv-TasNet and the clean utterance. Conv-TasNet significantly outperforms IRM ($p<1e-16$, t-test). (B): PESQ scores are higher for IRM compared to the Conv-TasNet ($p<1e-16$, t-test). Error bars indicate standard error (STE) (C): MOS versus PESQ for individual utterances. Each dot denotes one mixture utterance, separated using the IRM (blue) or Conv-TasNet (red). The subjective ratings of almost all utterances for Conv-TasNet are higher than their corresponding PESQ scores.}
	\label{fig:subjective}
\end{figure*}

Figure~\ref{fig:subjective} and table~\ref{tab:sub-comp} show the result of the human subjective quality test, where the MOS for Conv-TasNet is significantly higher than the MOS for the IRM ($p<1e-16$, t-test). In addition, the superior subjective quality of Conv-TasNet over IRM is consistent across most of the 25 test utterances as shown in figure~\ref{fig:subjective} (C). This observation shows that PESQ consistently underestimates MOS for Conv-TasNet separated utterances, which may be due to the dependence of PESQ on the magnitude spectrogram of speech \cite{rix2001perceptual} which could produce lower scores for time-domain approaches. 

\begin{table}[!htbp]
	\small
	\centering
	\caption{PESQ scores for the ideal T-F masks and Conv-TasNet on the entire WSJ0-2mix and WSJ0-3mix test sets.}
	\vspace{0.2cm}
	\label{tab:sub-masks}
	\begin{tabular}{c|cccc}
		\thline
		\multirow{2}{*}{\thead{Dataset}} & \multicolumn{4}{c}{\thead{PESQ}}\\
		& \thead{IRM} & \thead{IBM} & \thead{WFM} & \thead{Conv-TasNet}\\
		\thline
		WSJ0-2mix & \bf{3.74} & 3.33 & 3.70 & 3.24 \\
		WSJ0-3mix & \bf{3.52} & 2.91 & 3.45 & 2.61 \\
		\thline
	\end{tabular}
\end{table}

\begin{table}[!htbp]
	\small
	\centering
	\caption{Mean opinion score (MOS, N=40) and PESQ for the 25 selected utterances from the WSJ0-2mix test set.}
	\vspace{0.2cm}
	\label{tab:sub-comp}
	\begin{tabular}{c|c|c}
		\thline
		\thead{Method} & \thead{MOS} & \thead{PESQ} \\
		\thline
		\bf{Conv-TasNet-gLN} & \bf{4.03} & 3.22 \\
		IRM & 3.51 & \bf{3.74} \\
		\hline
		Clean & 4.23 & 4.5 \\
		\thline
	\end{tabular}
\end{table}

\subsection{Processing speed comparison}

Table~\ref{tab:speed} compares the processing speed of LSTM-TasNet and causal Conv-TasNet. The speed is evaluated as the average processing time for the systems to separate each frame in the mixtures, which we refer to as time per frame (TPF). TPF determines whether a system can be implemented in real time, which requires a TPF that is smaller than the frame length. 

For the CPU configuration, we tested the system with one processor on an Intel Core i7-5820K CPU. For the GPU configuration, we preloaded both the systems and the data to a Nvidia Titan Xp GPU. LSTM-TasNet with CPU configuration has a TPF close to its frame length (5 ms), which is only marginally acceptable in applications where only a slower CPU is available. Moreover, the processing in LSTM-TasNet is done sequentially, which means that the processing of each time frame must wait for the completion of the previous time frame, further increasing the total processing time of the entire utterance. Since Conv-TasNet decouples the processing of consecutive frames, the processing of subsequent frames does not have to wait until the completion of the current frame and allows the possibility of parallel computing. This process leads to a TPF that is 5 times smaller than the frame length (2 ms) in our CPU configuration. Therefore, even with slower CPUs, Conv-TasNet can still perform real-time separation.

\begin{table}[!htbp]
	\small
	\centering
	\caption{Processing time for causal LSTM-TasNet and Conv-TasNet. The speed is evaluated as the average time required to separate a frame (time per frame, TPF).}
	\vspace{0.2cm}
	\label{tab:speed}
	\begin{tabular}{c|c}
		\thline
		Method & CPU/GPU TPF (ms) \\
		\hline
		LSTM-TasNet & 4.3/0.2 \\
		Conv-TasNet-cLN & \bf{0.4}/\bf{0.02} \\
		\thline
	\end{tabular}
\end{table}

\subsection{Sensitivity of LSTM-TasNet to the mixture starting point}

\begin{figure}[!htp]
	\small
	\centering
	\includegraphics[width=\columnwidth]{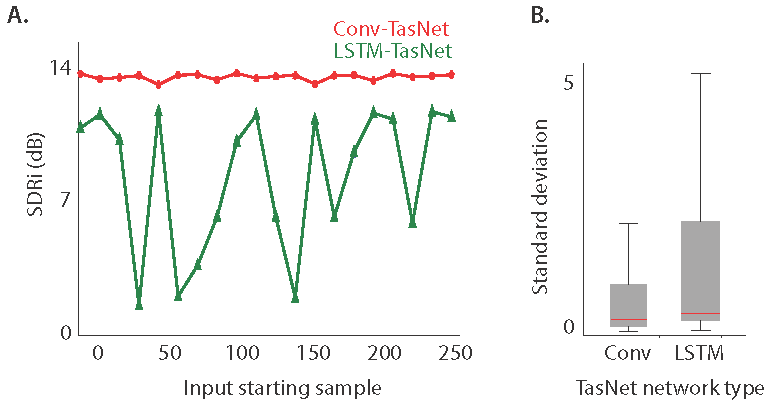}
	\caption{(A): SDRi of an example mixture separated using LSTM-TasNet and causal Conv-TasNet as a function of the starting point in the mixture. The performance of Conv-TasNet is considerably more consistent and insensitive to the start point. (B): Standard deviation of SDRi across all the mixtures in the WSJ0-2mix test set with varying starting points.}
	\label{fig:shift}
\end{figure}

Unlike language processing tasks where sentences have determined starting words, it is difficult to define a general starting sample or frame for speech separation and enhancement tasks. A robust audio processing system should therefore be insensitive to the starting point of the mixture. However, we empirically found that the performance of the causal LSTM-TasNet is very sensitive to the exact starting point of the mixture, which means that shifting the input mixture by several samples may adversely affect the separation accuracy. We systematically examined the robustness of LSTM-TasNet and causal Conv-TasNet to the starting point of the mixture by evaluating the separation accuracy for each mixture in the WSJ0-2mix test set with different sample shifts of the input. A shift of $s$ samples corresponds to starting the separation at sample $s$ instead of the first sample. Figure~\ref{fig:shift} (A) shows the performance of both systems on the same example mixture with different values of input shift. We observe that, unlike LSTM-TasNet, the causal Conv-TasNet performs consistently well for all shift values of the input mixture. We further tested the overall robustness for the entire test set by calculating the standard deviation of SDRi in each mixture with shifted mixture inputs similar to figure~\ref{fig:shift} (A). The box plots of all the mixtures in the WSJ0-2mix test set in figure~\ref{fig:shift} (B) show that causal Conv-TasNet performs consistently better across the entire test set, which confirms the robustness of Conv-TasNet to variations in the starting point of the mixture. One explanation for this inconsistency may be due to the sequential processing constraint in LSTM-TasNet which means that failures in previous frames can accumulate and affect the separation performance in all following frames, while the decoupled processing of consecutive frames in Conv-TasNet alleviates the effect of occasional error.

\subsection{Properties of the basis functions}
\label{sec:exp-basis}

\begin{figure}[!htbp]
	\small
	\centering
	\includegraphics[width=\columnwidth]{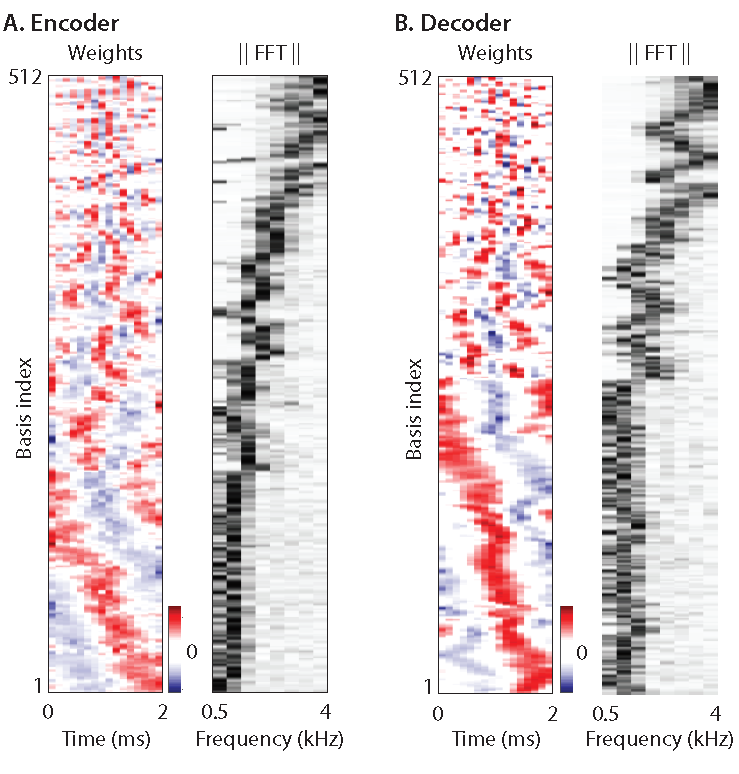}
	\caption{Visualization of encoder and decoder basis functions and the magnitudes of their FFTs. The basis functions are sorted based on their pairwise Euclidean similarity.}
	\label{fig:basis}
\end{figure}

One of the motivations for replacing the STFT representation of the mixture signal with the convolutional encoder in TasNet was to construct a representation of the audio that is optimized for speech separation. To shed light on the properties of the encoder and decoder representations, we examine the basis functions of the encoder and decoder (rows of the matrices $\vec{U}$ and $\vec{V}$). The basis functions are shown in figure~\ref{fig:basis} for the best noncausal Conv-TasNet, sorted in the same way as figure~\ref{fig:sample}. The magnitudes of the FFTs for each filter are also shown in the same order. As seen in the figure, the majority of the filters are tuned to lower frequencies. In addition, it shows that filters with the same frequency tuning express various phase values for that frequency. This observation can be seen by the circular shift of the low-frequency basis functions. This result suggests an important role for low-frequency features of speech such as pitch as well as explicit encoding of the phase information to achieve superior speech separation performance.  

\section{Discussion}
\label{sec:conclude}
In this paper, we introduced the fully-convolutional time-domain audio separation network (Conv-TasNet), a deep learning framework for time-domain speech separation. This framework addresses the shortcomings of speech separation in the STFT domain, including the decoupling of phase and magnitude, the suboptimal representation of the mixture audio for separation, and the high latency of calculating the STFT. The improvements are accomplished by replacing the STFT with a convolutional encoder-decoder architecture. The separation in Conv-TasNet is done using a temporal convolutional network (TCN) architecture together with a depthwise separable convolution operation to address the challenges of deep LSTM networks. Our evaluations showed that Conv-TasNet significantly outperforms STFT speech separation systems even when the ideal time-frequency masks for the target speakers are used. In addition, Conv-TasNet has a smaller model size and a shorter minimum latency, which makes it suitable for low-resource, low latency applications.

Unlike STFT which has a well-defined inverse transform that can perfectly reconstruct the input, best performance in the proposed model is achieved by an overcomplete linear convolutional encoder-decoder framework without guaranteeing the perfect reconstruction of the input. This observation motivates rethinking of autoencoder and overcompleteness in the source separation problem which may share similarities to the studies of overcomplete dictionary and sparse coding \cite{lee1999blind, zibulevsky2001blind}. Moreover, the analysis of the encoder/decoder basis functions in section~\ref{sec:exp-basis} revealed two interesting properties. First, most of the filters are tuned to low acoustic frequencies (more than 60\% tuned to frequencies below 1 kHz). This pattern of frequency representation, which we found using a data-driven method, roughly resembles the well-known mel-frequency scale \cite{imai1983cepstral} as well as the tonotopic organization of the frequencies in the mammalian auditory system \cite{romani1982tonotopic, pantev1989tonotopic}. In addition, the overexpression of lower frequencies may indicate the importance of accurate pitch tracking in speech separation, similar to what has been reported in human multitalker perception studies \cite{darwin2003effects}. In addition, we found that filters with the same frequency tuning explicitly express various phase information. In contrast, this information is implicit in the STFT operations, where the real and imaginary parts only represent symmetric (cosine) and asymmetric (sine) phases, respectively. This explicit encoding of signal phase values may be the key reason for the superior performance of TasNet over the STFT-based separation methods.  

The combination of high accuracy, short latency, and small model size makes Conv-TasNet a suitable choice for both offline and real-time, low-latency speech processing applications such as embedded systems and wearable hearing and telecommunication devices. Conv-TasNet can also serve as a front-end module for tandem systems in other audio processing tasks, such as multitalker speech recognition \cite{hershey2010super, weng2015deep, qian2017single, ochi2016multi} and speaker identification \cite{lei2014novel, mclaren2015advances}. On the other hand, several limitations of Conv-TasNet must be addressed before it can be actualized, including the long-term tracking of speakers and generalization to noisy and reverberant environments. Because Conv-TasNet uses a fixed temporal context length, the long-term tracking of an individual speaker may fail, particularly when there is a long pause in the mixture audio. In addition, the generalization of Conv-TasNet to noisy and reverberant conditions must be further tested \cite{luo2018real}, as time-domain approaches are more prone to temporal distortions which are particularly severe in reverberant acoustic environments. In such conditions, extending the Conv-TasNet framework to incorporate multiple input audio channels may prove advantageous when more than one microphone is available. Previous studies have shown the benefit of extending speech separation to multichannel inputs \cite{gannot2017consolidated, chen2017cracking, wang2018multi}, particularly in adverse acoustic conditions and when the number of interfering speakers is large (e.g., more than 3). 

In summary, Conv-TasNet represents a significant step toward the realization of speech separation algorithms and opens many future research directions that would further improve its accuracy, speed, and computational cost, which could eventually make automatic speech separation a common and necessary feature of every speech processing technology designed for real-world applications.

\iftrue
\section{Acknowledgments}
This work was funded by a grant from the National Institute of Health, NIDCD, DC014279; a National Science Foundation CAREER Award; and the Pew Charitable Trusts.
\fi
\vfill\pagebreak
\bibliographystyle{IEEEbib}
\bibliography{refs}

\end{document}